\documentclass[reprint,twocolumn,aps,prl,superscriptaddress]{revtex4-2}
\usepackage{mhchem}
\usepackage{float}
\usepackage[english]{babel}
\usepackage[utf8]{inputenc}
\usepackage{fancyhdr}
\usepackage{sidecap}
\usepackage{multirow}

\usepackage{graphicx}
\usepackage{tabularx}
\usepackage{braket}
\usepackage{tikz}
\usepackage{color}
\usepackage{amsmath}
\usepackage[colorlinks=True, linkcolor=blue, filecolor=magenta, urlcolor=blue,citecolor=blue]{hyperref}
\usepackage[nameinlink,capitalise]{cleveref}
\begin{document}
\title{Thermodynamic and transport properties of high-quality single crystals \\
of the altermagnet CrSb}

\author{Shubhankar Paul}
\email{shubhp@iitk.ac.in}
\affiliation{Department of Physics, Indian Institute of Technology Kanpur, Kanpur 208016, India}
\affiliation{Department of Electronic Science and Engineering, Graduate School of Engineering, Kyoto University, Kyoto 615-8510, Japan}
\affiliation{Toyota Riken–Kyoto University Research Center (TRiKUC), Kyoto 606-8501, Japan}
\author{Atsutoshi Ikeda}
\affiliation{Department of Electronic Science and Engineering, Graduate School of Engineering, Kyoto University, Kyoto 615-8510, Japan}
\author{Giordano Mattoni}
\affiliation{Toyota Riken–Kyoto University Research Center (TRiKUC), Kyoto 606-8501, Japan}
\author{Shingo Yonezawa}
\affiliation{Department of Electronic Science and Engineering, Graduate School of Engineering, Kyoto University, Kyoto 615-8510, Japan}
\author{Chanchal Sow}
\email{chanchal@iitk.ac.in}
\affiliation{Department of Physics, Indian Institute of Technology Kanpur, Kanpur 208016, India}

\date{\today}
\begin{abstract}
Altermagnetism (AM) is an emerging magnetic order unifying essential characteristics of ferromagnetic and antiferromagnetic states.
The CrSb has attracted significant interest owing to its large altermagnetic spin-splitting energy.
In this paper, we present the growth details of high-quality single crystals of CrSb using the self-flux method and investigate their physical properties.
We obtained large
(001) oriented hexagonal crystals, up to 2 × 2.5 × 1 mm$^3$ in size with residual resistivity ratio $\sim$ 11.
A pronounced positive magnetoresistance of up to 80\% is observed at 3.5 K.
Most strikingly, the room temperature specific heat value exceeds the Dulong-Petit limit, being attributed to a broad magnon contribution from the altermagnetic order of CrSb.
The specific heat fit reveals a magnon energy gap $\sim$ 16 $\pm$ 1 meV.
Further, ac susceptibility measurements demonstrate the absence of superconductivity down to 0.1 K.
These findings underscore CrSb as a viable altermagnet for room temperature magnonic and spintronic applications.

\end{abstract}
\maketitle

\section*{Introduction}
\begin{figure}[b]
   \begin{center}
   \includegraphics[scale=0.32]{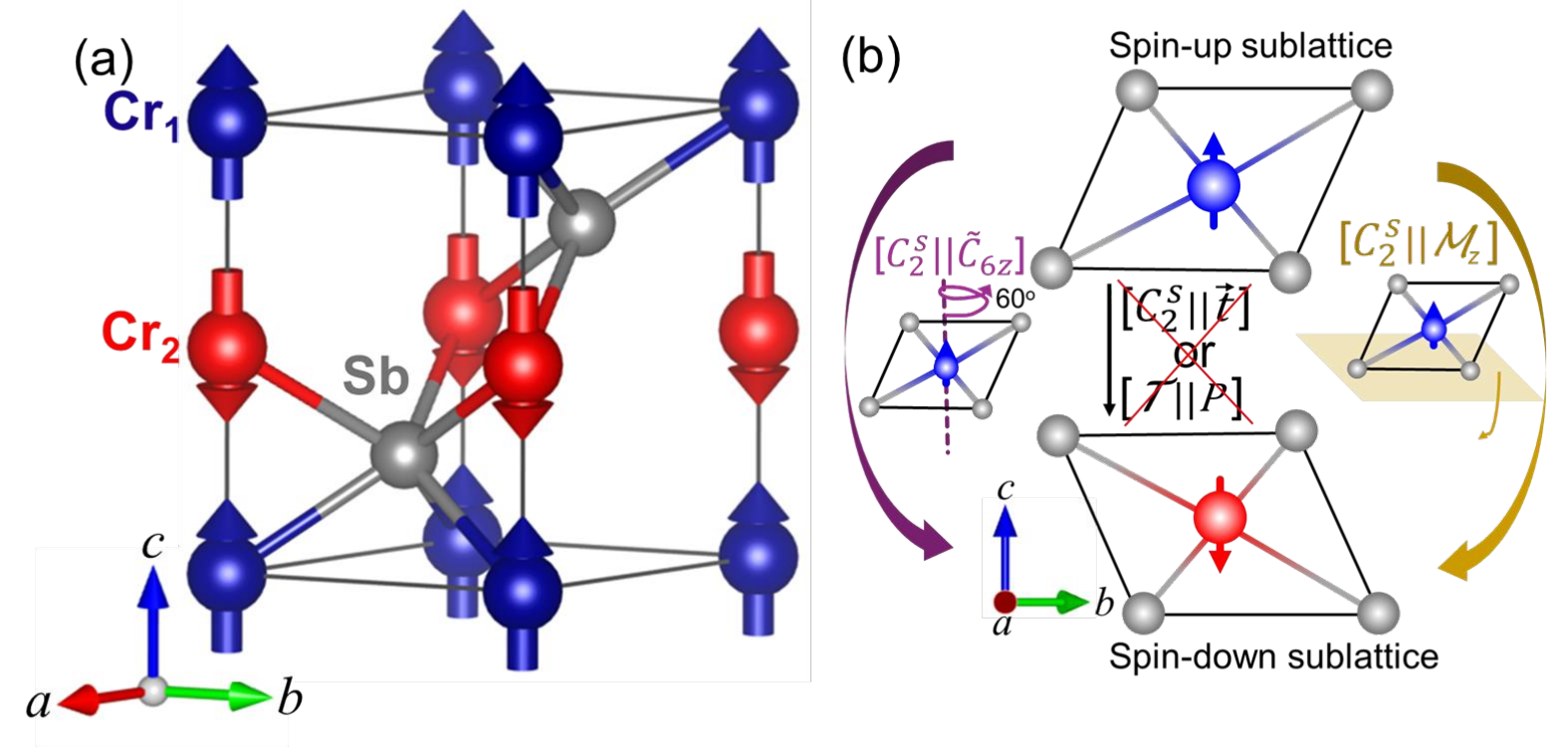}
   \caption{(a) Schamatic diagram of crystal and magnetic structures drawn by VESTA \cite{rodriguez1993recent}. Opposite-spin sublattices are denoted as Cr$_1$ (blue: spin-up) and Cr$_2$ (red: spin-down), with Sb atoms shown as grey spheres. (b) Schematic view of symmetry-driven sublattice interchange in CrSb. The two magnetic sublattices are related through either a screw operation accompanied by a half unit cell translation along the $c$ direction or a mirror operation with respect to a plane perpendicular to the $c$ axis.}
   \label {fig0}
   \end{center}
 \end{figure}

A new magnetic state termed altermagnetism (AM) has recently been proposed through the reclassification of antiferromagnets (AFMs) within the framework of spin groups \cite{vsmejkal2022beyond,vsmejkal2022emerging,mazin2022altermagnetism,jiang2024enumeration}.
Among the wide variety of altermagnets (AMs) predicted theoretically \cite{vsmejkal2022emerging}, RuO$_2$ \cite{paul2025growth,wu2025fermi, wvs6hqfv, karube2022observation, Kiefer2025}, and MnTe \cite{Fedchenko2024SciAdv} have been investigated most extensively. 
Although CrSb shows a remarkably high magnetic ordering temperature near 700 K \cite{hirone1956magnetic,abe1984magnetic} and substantial spin-splitting energy \cite{vsmejkal2022emerging,reimers2024direct}, it has drawn relatively less attention.
CrSb adopts a hexagonal NiAs-type crystal structure (space group: $P6_3/mmc$) \cite{hirone1956magnetic,reimers2024direct,abe1984magnetic,urata2024high,bai2025nonlinear,rai2025direction,peng2025scaling,aota2025epitaxial,bommanaboyena2025single} and exhibits a collinear antiferromagnetic ground state \cite{snow1952neutron,takei1963magnetic,singh2025chiral,zhang2025chiral,tas2025magnon}, with Cr moments oriented along the $c$-axis, ferromagnetically aligned within each hexagonal plane and antiferromagnetically coupled between neighboring planes, as shown in \cref{fig0}\textcolor{blue}{a}.
The altermagnetic order in CrSb can be well described within the framework of nonrelativistic spin-group symmetries.
A spin space group operation is defined as a combined transformation in spin and real space, defined as [$R_i^s||R_j$], where $R_i^s$ acts exclusively on the spin degrees of freedom (restricted to the identity $E$ and the spin-flip symmetry operation $C_2^s$) and $R_j$ denotes a spatial symmetry operation \cite{vsmejkal2022beyond,vsmejkal2022emerging,jiang2024enumeration,mazin2022altermagnetism,yang2025three}. The opposite-spin sublattices in altermagnet must be related by a nontrivial spatial symmetry—such as an n-fold rotation (screw) or mirror ($\mathcal{M}$) symmetry (glide) instead of translation ($\vec t$) or inversion ($P$).
The antiparallel Cr sublattices in CrSb are connected by the spin group symmetries $[{C}_2^s||\mathcal{M}_z]$ and $[C_2^s||\tilde{C}_{6z}]$, where $\tilde{C}_{6z}$ denotes a sixfold screw operation about the $z$ axis accompanied by a half-unit-cell translation ($\vec t$ = $(0,0,\frac{c}{2})$), as shown in \cref{fig0}\textcolor{blue}{b}.
Breaking of the spin group symmetries $[C_2^s||\vec t]$ and space-time reversal symmetry $[\mathcal{T}||{P}]$ in CrSb leads to momentum-dependent spin-splitting of the electronic bands \cite{vsmejkal2022beyond,vsmejkal2022emerging,jiang2024enumeration,mazin2022altermagnetism,yang2025three}.
This symmetry identifies CrSb as a bulk g-wave altermagnet \cite{vsmejkal2022emerging}, characterized by spin-split electronic bands that reverse polarity six times in the $k_x$–$k_y$ plane around $k_z$ axis. Such altermagnetic material retains key advantages of antiferromagnets—such as ultrafast spin dynamics and weak sensitivity to external magnetic fields—making it a promising platform for novel spin-to-charge conversion, spin-torque effects, and future spintronic or magneto-optical applications \cite{karube2022observation, Bai2022PRL, vsmejkal2022emerging,jungwirth2016antiferromagnetic,fukami2020antiferromagnetic}.
\begin{table*}
 	\caption{\mbox{Comparison of magnon characteristics in ferromagnet (FM), antiferromagnet (AFM) and altermagnet (AM).}}
 	\label{tab0}
 	\begin{tabular*}{1\textwidth}{@{\extracolsep{\fill}}cccccccl}
 		\hline     
 		\hline
Parameter &Isotropic& Anisotropic&Isotropic & Anisotropic  & Weak-SOC & Moderate/strong-\\
 &FM &  FM &  AFM  &AFM& AM&  SOC AM\\
\hline
\vspace{0mm}
    Low $k$ & quadratic &quadratic &linear &mixed&linear& mixed  \\
    dispersion &($\omega \propto \Delta+ D'k^2$)&($\omega \propto \Delta+ D'k^2$) & ($\omega \propto vk$) & ($\omega \propto \sqrt{\Delta+ v^2k^2}$)&$\omega_{(\uparrow)} (k) \neq \omega_{(\downarrow)} (k)$ &$\omega_{(\uparrow)} (k) \neq \omega_{(\downarrow)} (k)$\\
    \vspace{2mm}
    &$\Delta$ = 0  & $\Delta$ $\neq$ 0& $\omega_{(\uparrow)} (k)= \omega_{(\downarrow)} (k)$&$\Delta$ $\neq$ 0 & $\omega(k\rightarrow0)\rightarrow0$ &$\omega(k\rightarrow0)=\Delta_{\uparrow,\downarrow}$\\
\vspace{2mm}
Spin-wave branch &  non-degenerate& non-degenerate &  degenerate & non-degenerate, split &  non-degenerate & non-degenerate, split\\
\vspace{2mm}
spin polarization & Yes&Yes & No&Yes, $k$-dependent &Yes, $k$-dependent&Yes, $k$-dependent \\
\vspace{2mm}
$C_\mathrm{mag}$ & $\propto T^\mathrm{3/2}$&$\propto \  T^{\frac{1}{2}}\  \mathrm{exp}(-\frac{\Delta}{T}) $ & $ \propto T^\mathrm{3}$& $\propto \  T^{\frac{1}{2}}\  \mathrm{exp}(-\frac{\Delta}{T})$&power law& $\propto \  T^{\frac{1}{2}}\  \mathrm{exp}(-\frac{\Delta}{T})$\\
\vspace{2mm}
Magnon gap ($\Delta$) & gapless& gapped & gapless&gapped & gapless & gapped\\
\hline
\hline
\end{tabular*}
\end{table*}

The electronic and spin-transport properties of altermagnets have attracted significant attention, but their thermodynamic signatures, particularly the specific heat, remain comparatively less explored.
Specific heat would provide a direct thermodynamic measure of low-energy electronic and magnetic excitations of the altermagnetic state.
In conventional AFMs with linear spin-wave dispersion, the low-temperature specific heat is typically governed by gapless magnons with a characteristic power law dependence ($C_\mathrm{mag} \propto T^{d}$, where $d$ = 1, 2, 3 for a 1-D (linear), 2-D (layer), and 3-D (spatial) antiferromagnet, respectively) \cite{corticelli2022spin,joshua1998magnon,vsmejkal2023chiral}, whereas isotropic ferromagnets (FMs) with quadratic spin wave dispersion, the magnetic contribution typically follows a $T^{3/2}$ dependence \cite{vsmejkal2023chiral, varshney2004analysis}.
In altermagnets, symmetry induces anisotropic and spin-split altermagnon modes without net magnetization \cite{vsmejkal2023chiral,liu2024chiral,singh2025chiral,zhang2025chiral,tas2025magnon}.
A comparative overview of magnon dispersions and the anticipated magnetic specific-heat behaviour in FMs, AFMs, and AMs systems is summarised in \cref{tab0}.
Recent neutron diffraction studies on CrSb demonstrate that altermagnetic order lifts altermagnon spin degeneracy and opens finite gaps in the magnon spectrum \cite{singh2025chiral,zhang2025chiral,tas2025magnon}.
Thus, specific heat measurements provide a powerful probe of symmetry-driven magnetic excitations in altermagnets.
It has been reported that the specific heat of CrSb from room temperature to 900 K exhibits a pronounced rise near 685 K, exhibiting an antiferromagnetic transition \cite{abe1984magnetic}.
However, a detailed study of the low-temperature specific heat in CrSb single crystals is still lacking.
Moreover, there are no detailed reports on the growth of high-quality CrSb single crystals.
Most existing reports use Sn flux and yield needle-like crystals with a cross-section of about 0.3 mm \cite{urata2024high,rai2025direction,bai2025nonlinear}.

In this work, we report our new approach to the growth of high-quality CrSb single crystals using the self-flux method and investigate their low-temperature thermodynamic and transport properties.
We revealed that our Sn free growth technique significantly enhances both the crystal size and quality.
Our large crystals enable us to investigate thermodynamic properties using one crystal, allowing more accurate measurements free from complications arising from the use of multiple crystals.
Comprehensive measurements of specific heat uncover the magnetic contributions over a wide temperature range, which play a crucial role in the temperature dependence of specific heat.
This provides thermodynamic support for the existence of the gapped magnons from the altermagnetic order.
This result shows that CrSb gives rise to robust gapped magnon modes that remain stable far above room temperature, making it a promising candidate for room temperature magnonic and spin-transport applications.
In addition, it is important to examine the existence of superconductivity in CrSb, since non-stoichiometric polycrystalline CrSb$_{1+\delta}$ is reported to show superconductivity below  $T_\mathrm{c}$ = 8.8 K \cite{dahal2017possible}.
We report that superconductivity in our CrSb crystal is absent down to \mbox{0.1 K}.
 \begin{table*}
    \small
    \caption{Conditions of the growth of CrSb crystals. Temperature is monitored by the system's thermocouple.}
    \label{tab1}
    \begin{tabularx}{1\textwidth}{clcccclclX}
 	\hline
        \hline
     Run  & Cr:Sb  & Growth  & Centrifuge  & Cooling  &  & &  Typical crystal\\ [-1pt]
     number& ratio & temp. ($^{\circ}$C) & temp ($^{\circ}$C) & rate ($^\mathrm{o}$/hrs) & Remarks  & &size (mm$^3$)\\[+2pt]
        \hline 
     1 & 30:70 & 1110 & 670 &  2   & CrSb$_2$ crystals & &1$\times$1$\times$0.5
\\ [+2pt]
     2 & 45:55 & 1000 &  900 & 2   & Tiny CrSb crystals & &0.1$\times$0.1$\times$0.1 \\ [+2pt]
     3 & 45:55 & 1000 &  670 & 2   & Large hexagonal-shape crystals and lamellar crystals & & 1.5-2$\times$2.5$\times$1  \\ [+2pt]
     4 & 45:55 & 1000 &  670 & 2  & Large hexagonal-shape crystals and lamellar crystals & & 1-2.5$\times$2$\times$1\\  [+2pt]
     5 & 45:55 & 1000 & 600 & 2  & Flux-colated crystals& & 3$\times$2.5$\times$1.5 \\ [+2pt]
     6 & 45:55 & 1110 & 670 &  1.5   & Tiny CrSb crystals & & 0.1-0.5$\times$0.1$\times$0.2\\
        \hline
        \hline
    \end{tabularx}
\end{table*}
\section*{ Single crystal growth}
\subsection*{Growth technique}
\begin{figure}[b]
   \begin{center}
   \includegraphics[scale=0.45]{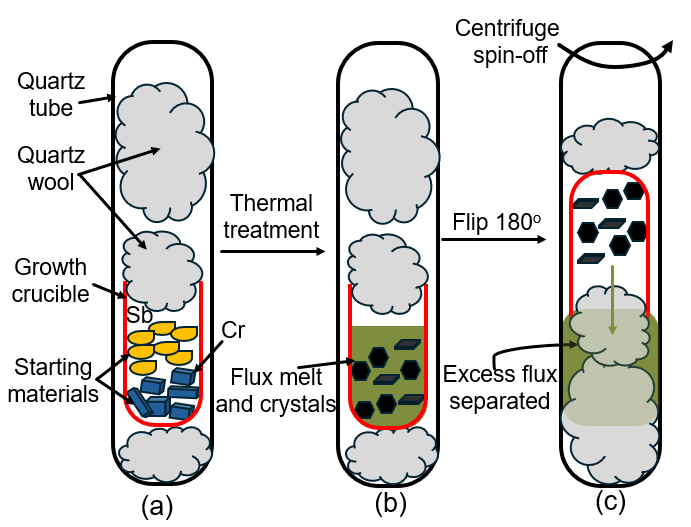}
   \caption{Schematic description of the CrSb single crystal growth process based on the self-flux method. (a) Raw materials (Cr and Sb) are placed in a crucible inside a quartz ampoule for thermal treatment. (b) Upon heating, the flux melts, and CrSb crystals form. (c) The quartz ampoule is inverted by 180$^{\circ}$ and centrifuged to separate the residual flux from the CrSb crystals.}
   \label {fig1}
   \end{center}
 \end{figure}

We employed the self-flux method using high-purity Cr chunks (Thermo Fisher Scientific, 99.995\%) and Sb shots (Thermo Fisher Scientific, 99.999\%) as starting materials.
The materials were loaded into an alumina crucible of dimensions 25 mm (length) $\times$ 10 mm (inner diameter) that is inert to the melt.
We put the material with a higher melting temperature at the bottom.
As the lower-melting-point material (Sb) liquefy, they flow over the higher-melting-point one (Cr), gradually dissolving them into the melt.
A plug of quartz wool was used as a cap to serve as a filter during the flux separation process. The crucible was placed inside a quartz ampoule, slightly raised above the ampoule base using small pieces of quartz wool to prevent cracking due to thermal expansion mismatch between the crucible and the quartz tube.
Additional quartz wool was placed above the crucible as shown in ~\cref{fig1}. This arrangement helps protect the quartz tube from damage during the high-speed centrifugation step. After assembling the setup, the quartz tube was purged several times with argon and then sealed under high vacuum (10$^{-5}$ mbar). Finally, the sealed tube was placed vertically inside a box furnace for thermal treatment.
\begin{figure*}[t]
   \begin{center}
   \includegraphics[scale=0.55]{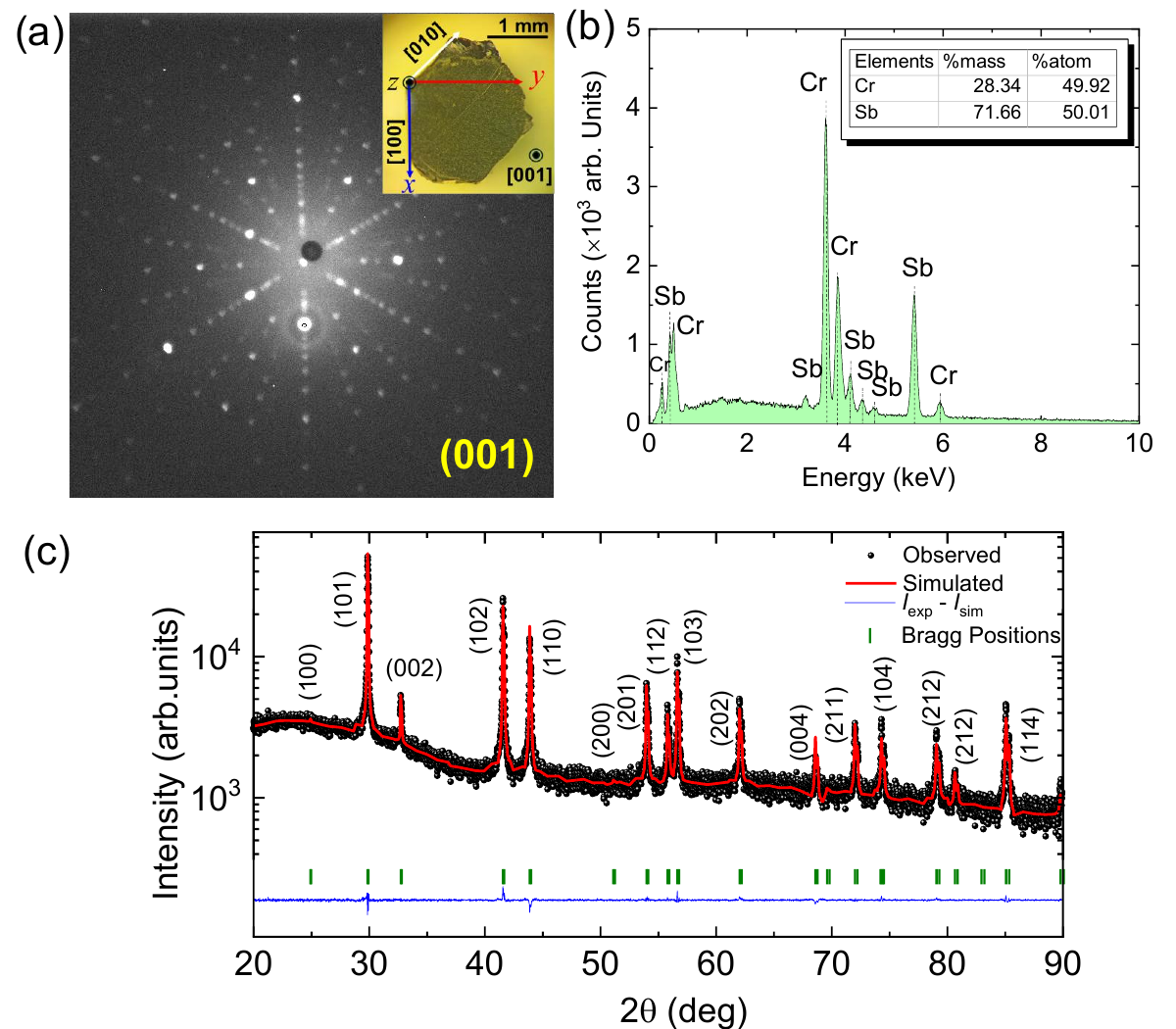}
   \caption{(a) Laue diffraction pattern of a CrSb crystal, showing a diffraction pattern characteristic of the (001) plane. Inset: optical image of the CrSb single crystal. (b) Result of energy dispersive spectroscopy (EDS) of the same crystal indicating a stoichiometric CrSb composition. (c) Powder XRD pattern of crushed CrSb single crystals (black points), shown together with the simulated profile (red curve). The difference between the experimental and simulated intensities is also shown with the blue curve. The green vertical lines represent the calculated Bragg reflection positions. The near coincidence of the two green bars arises from the $K_{\alpha_1}$ and $K_{\alpha_2}$ characteristic X-ray emission lines.}
   \label {fig2}
   \end{center}
 \end{figure*}

\subsection*{Growth conditions}
We observed a systematic variation in crystal formation as a function of growth temperature and stoichiometric ratio of the starting materials. The optimal molar ratio of Cr to Sb was determined to be 45:55. The sealed tube was placed inside a furnace, where the temperature was raised to 1110$^{\circ}$C at a rate of 200$^{\circ}$C/h and held for 36 hours.
After this, the sample was slowly cooled at a rate of 2$^{\circ}$C/h. At 900$^{\circ}$C, the tube was quickly taken out from the furnace, and inverted by 180$^{\circ}$ into a centrifuge cup, and rapidly spun. This process drives the molten flux through the quartz wool, effectively separating it from the crystals, which remain in the crucible. The schematic of the growth process is shown in \cref{fig1}.
We observed very tiny crystals inside the crucible.
In the subsequent trial, we reduced the centrifugation temperature to 670$^{\circ}$C and quenched the tube in ice water after centrifugation.
Rapid quenching makes the flux more brittle and easier to separate the residual flux mechanically from the crystals.
This resulted in the formation of well-defined hexagonal crystals, typically around 2$\times$2.5$\times$1 mm$^3$ in size.
We also performed centrifugation at 600$^{\circ}$C; although crystals were obtained, their separation from the residual flux was more challenging.
This is likely due to the solidification of Sb, which occurs at 630$^{\circ}$C. Below this temperature, solidified Sb tends to coat the crystal surfaces, hindering clean separation. Therefore, centrifuging above 670$^{\circ}$C is essential; otherwise, the rapid temperature drop from furnace to centrifuge can adversely affect the flux removal process. The growth conditions and their outcomes are summarized in \cref{tab1}.

\section*{Results and discussion}
\subsection*{Crystal characterization}
Inset of \cref{fig2}\textcolor{blue}{{a}} shows an optical image of a flat, hexagonal-shaped CrSb crystal with typical dimensions of approximately 2$\times$2.5$\times$1 mm$^3$. 
The crystallographic orientation was determined using Laue backscattering photos recorded using a RIGAKU RASCO-BLII system, as shown in \cref{fig2}\textcolor{blue}{a}, where the diffraction pattern indicates that the broad facet corresponds to the (001) plane.

Elemental analysis was conducted on both the as-grown surface and a freshly cleaved surface using an electron microscope equipped with tungsten filaments (W-SEM, JSM-6010LA; JEOL) to determine the precise stoichiometry of Cr and Sb. As presented in \cref{fig2}\textcolor{blue}{b}, the analysis for both surfaces shows a Cr:Sb atomic ratio of 1:1.
The phase purity and crystal structure were further examined through powder X-ray diffraction (XRD) on crushed crystals, using a desktop XRD spectrometer (RIGAKU MiniFlex 600-C) with Cu-$K_\alpha$ radiation. The diffraction pattern collected down to 2$\theta$ = 3$^{\circ}$ reveals no impurity phases (\cref{fig2}\textcolor{blue}{c}). 
We carried out Rietveld refinement of the diffraction profile using the FullProf suite \cite{rodriguez1993recent} to obtain the lattice parameters $a$ = 4.12 \AA\ and $c$ = 5.47 \AA, consistent with previously reported values of $a$ = 4.10--4.12 \AA\ and $c$ = 5.45--5.47\AA \ \cite{dahal2017possible,abe1984magnetic,urata2024high,bai2025nonlinear,urata2024high,reimers2024direct}.
\begin{figure*}
   \begin{center}
      \includegraphics[scale=0.5]{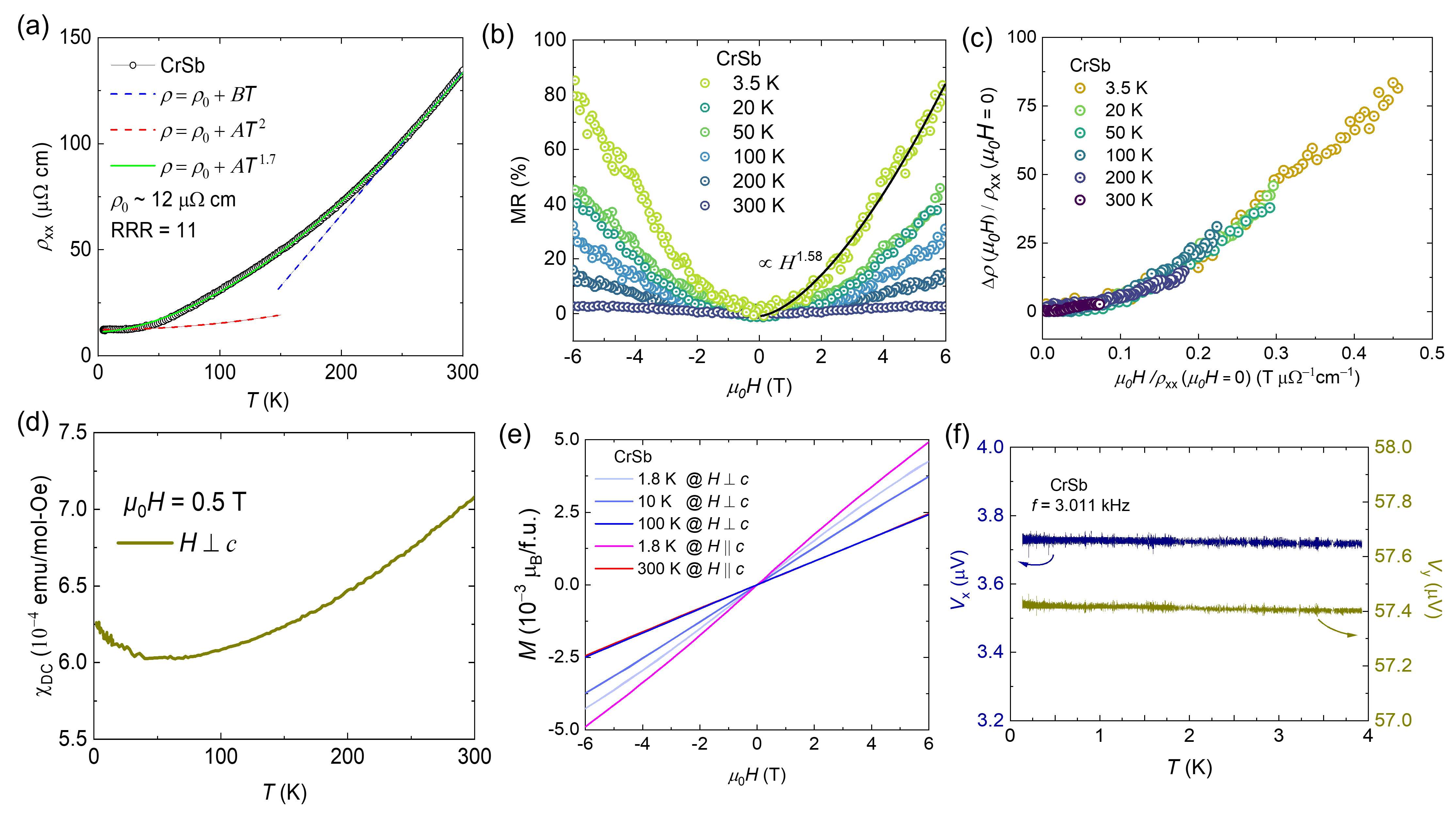}
     \caption{(a) Temperature dependence of the zero-field longitudinal resistivity. (b) MR at various temperatures ranging in 3.5$-$300 K for $H \parallel c$ and $I\parallel x$. (c) Kohler’s plots of the MR at different temperatures.
     (d) Temperature dependence of the DC magnetic susceptibility (1.8$-$300 K) for $H\perp$ $c$. (e) Field-dependent magnetization measured at various temperatures for $H \parallel$ $c$ and $H\perp$ $c$. (f) Susceptometer signal, which is proportional to the AC susceptibility, for a single crystal of CrSb. This measurement was performed using an AC excitation current of 0.50 mA rms ($\sim$ 17 $\mu$T rms) at a frequency of 3.011 kHz.}
    \label{fig4}
   \end{center}
\end{figure*}

\subsection*{Resistivity}
Figure~\ref{fig4}\textcolor{blue}{a} shows the temperature-dependent longitudinal resistivity measured from 1.8 K to 300 K using the AC four-probe technique \cite{paul2025growth,paul2026multi}.
Resistivity measurements were performed using a custom-made transport probe mounted on a Quantum Design MPMS-XL system.
We defined the $x$ axis as along one of the $a$ axis, the $y$ perpendicular to $x$ within the $ab$-plane, and the $z$ axis as along the $c$-axis, as illustrated in the inset of \cref{fig2}\textcolor{blue}{a}.
The electric current was applied along the $x$ axis.
The resistivity shows a monotonic increase with temperature, indicating metallic conduction.
In zero magnetic field, the longitudinal resistivity $\rho_{xx}$($T$) decreases from 135 $\mu\Omega$ cm at 300 K to 12 $\mu\Omega$ cm at 4 K, corresponding to a residual resistivity ratio (RRR: $\rho_{300\ \mathrm{K}}/\rho_{4\ \mathrm{K}}$) of about 11. This value is higher than those reported in recent studies \cite{urata2024high,bai2025nonlinear,rai2025direction,li2025large,bommanaboyena2025single,aota2025epitaxial}, indicating high crystal quality of our samples.
At high temperatures, electron–phonon scattering predominates, resulting in a linear dependence on $T$, while at low temperatures, electron-electron interactions take over, leading to a $T^2$ dependence. 
The blue and red dashed curves in \cref{fig4}\textcolor{blue}{a} represent fits to the data using $\rho = \rho_0 + BT$, and $\rho = \rho_0 + AT^2$, respectively. The overall temperature dependence in the entire range is well described by the empirical relation $\rho = \rho_0 + AT^{1.7}$. 
\begin{table}[b]
    \caption{Variation of crystal quality and MR at low temperature (below 10 K) for single crystals and thin films of CrSb.}
    \label{MR}
    \centering
    \begin{tabular}{cccc} 
 	\hline
     \hline \\
     RRR  & MR (\%)   & Thickness/size  & Ref.\\
        &at 6 T  & & \\ 
        \hline
     1.7 & 0.1&  Thin film (25 nm)& \cite{bommanaboyena2025single}\\ 
     4.3 & 1.25 &  Thin film (30 nm)& \cite{aota2025epitaxial}\\ 
          5 &15 & 1$\times$1$\times$0.3 mm$^3$& \cite{bai2025nonlinear}\\ 
   7.3& 16 & 3$\times$0.5$\times$ 0.1 mm$^3$& \cite{thadathil2026electrical} \\
         8.9 & 45  & 2$\times$2$\times$0.05 mm$^3$ & \cite{peng2025scaling}\\ 
    11 & 80 & 2$\times$ 2.5 $\times$1 mm$^3$&This work\\
    \hline
    \end{tabular}
\end{table}
\begin{figure*}
   \begin{center}
      \includegraphics[scale=0.53]{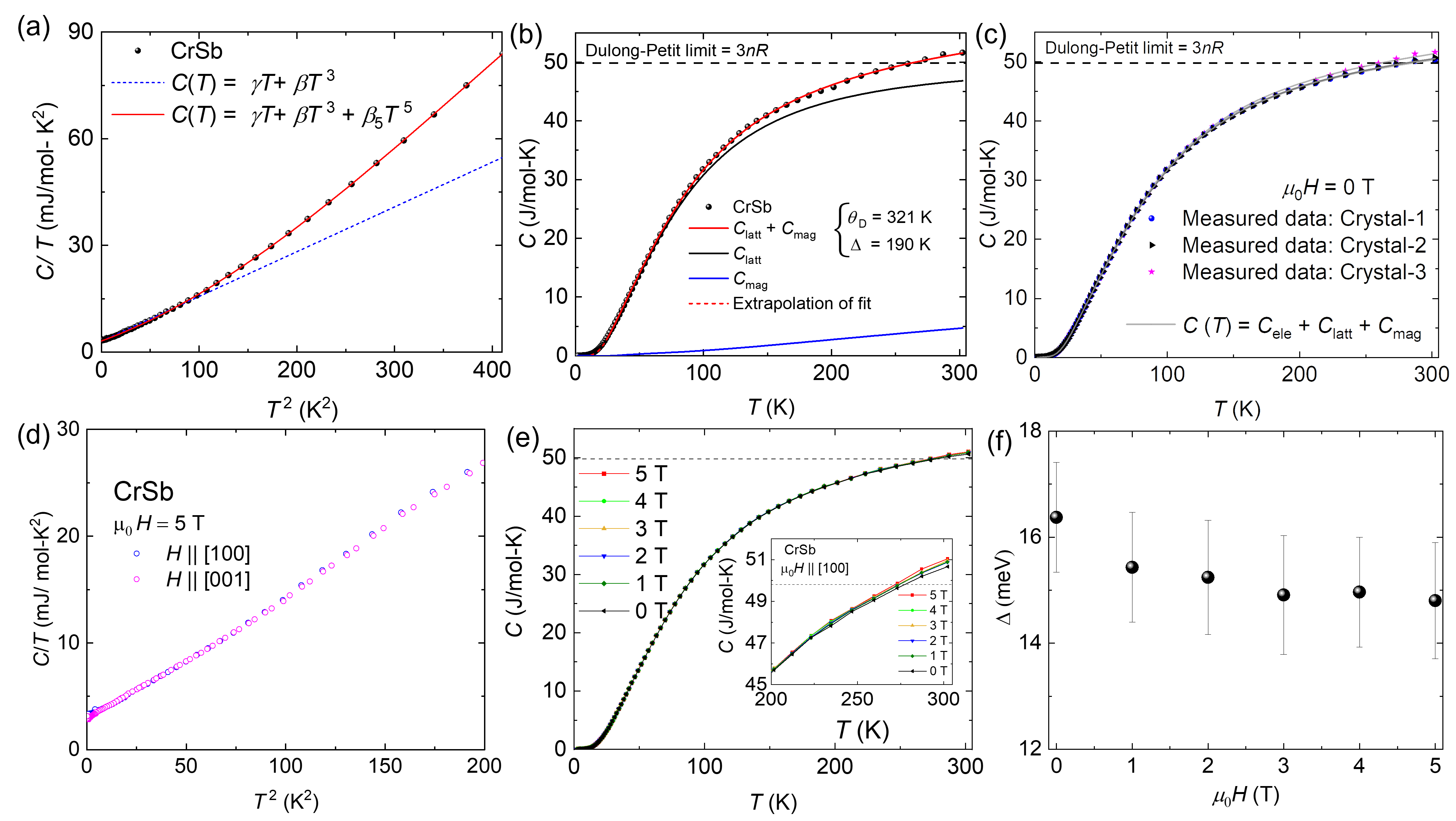}
       \caption{(a) $C$/T plotted against $T^2$ from 0.45 to 20 K. Data is fitted with \cref{eq2}. We also show the result of fitting with $C=\gamma T +\beta T^3$ below 8 K (blue broken curve). (b) Temperature-dependent specific heat of a CrSb crystal measured between 1.8 K and 300 K under zero magnetic field. The black dashed line corresponds to the Dulong Petit limit. The red solid line represents the fit obtained over the temperature range 25$–$300 K using \cref{eq7a}. The black and blue solid curves indicate the individually extracted lattice and magnon contributions, respectively. (c) Specific heat results from several samples. The solid curves are the result of fitting with \cref{eq7a}. The specific heat of RuO$_2$, expressed as ($C/T$), is plotted as a function of $T^{2}$ for temperatures between 0.5 and 10 K under a 5 T magnetic field applied along the [001] and [100] crystallographic directions. The data for the different orientations overlap within experimental accuracy, indicating isotropic behaviour. (e) Field and temperature dependence of the specific heat of CrSb. The inset shows the zoom-in view of the $C(T, H)$. (f) Variation of $\Delta$ with field.}
    \label {fig5}
   \end{center}
\end{figure*}

Figure \ref{fig4}\textcolor{blue}{b} shows the magnetic field-dependent magnetoresistance (MR =$ \frac{\rho_{xx}(H)-\rho_{xx} (\mu_0H=0)}{\rho_{xx}(\mu_0H=0)})$ in the temperature range 3.5--300 K.
A pronounced positive MR of nearly 80\% at 6 T and 3.5 K was observed, higher than values commonly reported for antiferromagnetic metals \cite{volny2020electrical,bodnar2020magnetoresistance,peng2025scaling} and those in recent studies on CrSb single crystals \cite{urata2024high,bai2025nonlinear,rai2025direction,peng2025scaling,thadathil2026electrical}, indicating the superior quality of the present crystal.
We compare our MR results with recent reports on CrSb single crystals \cite{bai2025nonlinear,peng2025scaling,thadathil2026electrical} and thin films \cite{aota2025epitaxial,bommanaboyena2025single}  (see \cref{MR}). Clearly, higher-RRR samples exhibit larger MR. This supports the attribution of higher MR in our samples to the improved crystal quality.
As the temperature increases, the MR gradually decreases in magnitude yet remains positive in the entire field range.
As shown by solid curve in \cref{fig4}\textcolor{blue}{b}, we found that MR varies with $B^{1.58}$, consistent with previous reports \cite{bai2025nonlinear, urata2024high}.
According to classic Kohler’s rule, magnetoresistance measured at different temperatures is expected to merge onto a universal curve when expressed as a function of the scaled field ($\mu_0H$/$\rho_{xx} (\mu_0H=0)$). As shown in \cref{fig4}\textcolor{blue}{c}, our MR data collapse onto a single curve, demonstrating strong adherence to Kohler’s rule. This behavior suggests that the magnetoresistance is primarily governed by the orbital motion of charge carriers with a common scattering time in a magnetic field \cite{bai2025nonlinear,urata2024high,rai2025direction}.

\subsection*{Magnetization}
The temperature-dependent DC susceptibility ($M/H$) of a CrSb single crystal was measured with the magnetic field applied perpendicular to the crystallographic $c$-axis ($H\bot c$), as presented in \cref{fig4}\textcolor{blue}{d}.
Magnetic properties were measured with a SQUID magnetometer (MPMS-XL, Quantum Design).
Néel temperature is expected to be well above room temperature, beyond the measured range of 1.8-300 K. Magnetization was found to decrease gradually upon cooling, consistent with the behavior expected for an antiferromagnetic material.

Figure \ref{fig4}\textcolor{blue}{e} presents the magnetization as a function of magnetic field at 1.8-300 K along $H\bot c$ and $H\parallel c$. 
The magnetization increases linearly with the applied magnetic field and shows no signs of magnetic hysteresis, spin-flop, or metamagnetic transitions. Even at 6 T, magnetization remains relatively small, approximately 10$^{-3}$ $\mu_\mathrm{B}$ per Cr atom, notably lower than earlier neutron diffraction studies \cite{abe1984magnetic,yuan2020magnetic}.
Moreover, the magnetization remains far from saturation even at 6 T, consistent with the typical behavior of antiferromagnetic materials.
The $M-H$ curve at 1.8 K is showing a weak non-linear trend, suggesting that the spins in the antiferromagnetic state are strongly coupled, allowing only a slight canting in response to the applied magnetic field \cite{urata2024high}.

\subsection{AC susceptibility down to 0.1 K}
In order to examine the superconductivity in CrSb, the AC susceptibility of the sample was measured using the mutual inductance technique with a lock-in amplifier, integrated into a Quantum Design Physical Property Measurement System (PPMS) \cite{yonezawa2015compact}.
The susceptometer is compatible with both the conventional PPMS sample chamber and the adiabatic demagnetization refrigerator (ADR) module. 
The ADR, employing a paramagnetic salt (chromium alum), allows rapid cooling from room temperature to below 0.1 K within two hours.
The susceptometer coil is made of an outer primary coil and inner secondary coils, detailed configuration outlined in \cite{yonezawa2015compact}.
The output signal from the susceptometer was detected using a lock-in amplifier (SR830, Stanford Research Systems), which enables measurement of both the in-phase ($V_x$) and out-of-phase ($V_y$) components of the AC voltage at a frequency of 3.011 kHz.
Figure \ref{fig4}\textcolor{blue}{f} shows susceptometer response in the range 0.1 K to 4.0 K measured using the ADR option. 
Our results show that there is no superconductivity in a stoichiometric CrSb single crystals down to 0.1 K.

\subsection*{Specific heat capacity}
Figure \ref{fig5}\textcolor{blue}{a} presents the temperature-dependent specific heat of a CrSb single crystal measured under zero magnetic field, with data collected in logarithmic temperature intervals from 0.45 to 20 K.
The specific heat measurements were carried out using a commercial calorimetry setup (Quantum Design PPMS) equipped with a $^3$He refrigeration option. 
We do not observe any phase transition down to 0.45 K.
The observed specific heat primarily consists of two contributions and is given by 
\begin{equation}
\label{eq1}
C(T)=  C_{\mathrm{ele}}(T) +C_{\mathrm{latt}}(T).
\end{equation}
The first term corresponds to the electronic specific heat associated with conduction electrons ($C_\mathrm{ele}$), expressed as $\gamma T$, where the Sommerfeld coefficient $\gamma = {\pi^2 N_\mathrm{A} k_\mathrm{B}^2\ D} (\epsilon_\mathrm{F})/{3}$, with $k_\mathrm{B}$ as Boltzmann’s constant and $D(\epsilon_\mathrm{F}$) the density of states (DOS) at the Fermi energy for both spin directions.
The second term represents lattice-specific heat arising from phonons ($C_\mathrm{latt}$).
We extracted the electronic contribution to the specific heat ($\gamma$) by plotting $C/T$ against $T^{2}$ in the 0.45--20 K range, as shown in \cref{fig5}\textcolor{blue}{a}.
The low-temperature data were fitted using a standard model with linear electronic and cubic lattice terms (blue broken curve), but deviated possibly due to higher-order phonon effects above $\sim$8 K.
As shown by the red curve, the low-temperature specific heat is well fit up to 20 K by:
\begin{equation}
  \label{eq2}
C(T) = \gamma T + \beta  T^3+ \beta_5 T^5,
\end{equation}
where $\beta$ is associated to the Debye temperature $\theta_\mathrm{D}$ as \cite{paul2025nonanalytic}
\begin{equation}
     \label{eq3}
   \beta = \dfrac{12\pi^4 n k_\mathrm{B} N_\mathrm{A}}{5\theta_\mathrm{D}^3},
\end{equation}
where $n$ = 2 corresponds to the total number of atoms in a formula unit (f.u.) for CrSb.
Fitting the zero-field data using \cref{eq2} gives $\gamma$ = 4.0 $\pm$ 0.08 mJ mol$^{-1}$ K$^{-2}$ and a phonon coefficient $\beta$ = 0.12 $\pm$ 0.02 mJ mol$^{-1}$ K$^{-4}$, corresponding to a Debye temperature $\theta_\mathrm{D}$ is 318 K.
We compare the obtained $\gamma$ with the electronic density of states (DOS) at the Fermi level, $D(\varepsilon_\mathrm {F})$.
Using $\gamma$, we estimate $D(\varepsilon_\mathrm {F})$ = 1.7 states/eV/f.u. = 23.1 states/Ry/f.u., consistent with DOS of about 1.6-1.7 states/eV/f.u.= 21.8 states/Ry/f.u. obtained from band-structure calculations without spin–orbit coupling \cite{kahal2007magnetic}.
We expect that spin-orbit interaction plays a minor role in DOS since it has been reported that band-structure calculations with and without SOC exhibit subtle differences in band structure near the Fermi level \cite{ding2024large,yang2025three}.

Now, we examine the high-temperature specific heat data (25 K to 300 K), presented in \cref{fig5}\textcolor{blue}{b}.
We found that the specific heat exceeds the Dulong–Petit limit (C$_V$ = 3$nR$ = 49.8 J mol$^{-1}$ K$^{-1}$, where $R$ is the gas constant) near room temperature.
This behaviour cannot be explained by only lattice and electronic heat capacity contributions.
To explain the specific heat exceeding the Dulong–Petit limit, it is necessary to consider the magnetic contribution.
Thus, the total specific heat can be expressed as: 
\begin{equation}
  \label{eq7a}
C(T)= \gamma T+ C_{\mathrm {V({Debye})}}(T) +C_\mathrm{{mag}}(T), 
\end{equation}
where $C_{V(\mathrm{Debye})}$$(T)$ is the specific heat at constant volume formulated as
\begin{equation}
  \label{eq4}
C_{\mathrm {V({Debye})}}(T) = 9nR \left(\frac{T}{\theta_\mathrm{D}}\right)^3 \int_{0}^{\theta_\mathrm{D}/T}\frac{x^4}{(e^x-1)^2} dx.
\end{equation}
For $C_\mathrm{{mag}}(T)$, the magnetic specific heat due to magnetic excitation, we attempted to use various models listed in \cref{tab0} and found that the gapless models cannot fit the experimental data.
Thus, we used the gapped magnon model, namely
\begin{equation}
\label{eq5}
C_\mathrm{{mag}}(T)= a_1T^{\frac{1}{2}}\mathrm{exp}(-\frac{\Delta}{T}),
\end{equation}
\noindent which accounts for the specific heat of magnons with a moderate energy gap $\Delta$.
The red solid line in \cref{fig5}\textcolor{blue}{b} shows the best fit over 25–300 K using \cref{eq7a}, while the black and blue solid curves represent the extracted lattice and magnon contributions, respectively.
At room temperature, the zero-field $C_\mathrm{mag}$ of CrSb is about 3.2 J/mol-K, consistent with earlier report \cite{abe1984magnetic}.
Within the considered temperature range, the best fit yields a Debye temperature of about 321 $\pm$ 5 K, in good agreement with the low-temperature estimate and the magnon energy gap of $\Delta$ = 190 $\pm$10 K ($\sim$ 16 $\pm$1 meV), consistent with inelastic neutron-scattering measurements \cite{radhakrishna1996inelastic,singh2025chiral}.
The corresponding fitted parameters for all samples are summarized in \cref{tab2}.
All samples exhibit similar values of $\theta_\mathrm{D}$ and $\Delta$, showing the high reproducibility of our investigation.
Among the gapped models, since the expected specific heat behaviour ($C_\mathrm{{mag}}(T) \propto T^{\frac{1}{2}}\mathrm{exp}(-\frac{\Delta}{T})$) is identical for all of anisotropic FM, AFM, and AM, it is difficult to distinguish the types of magnetic order only from the temperature dependence of specific heat. Nevertheless, the deduced gap size is in good agreement with previous experimental and theoretical studies on CrSb. This fact provides strong thermodynamic support for the altermagnetic order in CrSb.
\begin{table}
    \caption{Extracted parameters from the specific heat data of CrSb using the combination of lattice, electronic, and magnetic contributions. The value of $\theta_{\mathrm D}$ in this table was evaluated from high-temperature fitting using \cref{eq7a}.}
    \label{tab2}
    \centering
    \begin{tabular}{cccc} 
 	\hline
    \hline
     Parameters & Crystal-1 & Crystal-2  & Crystal-3 \\
     \hline
     mass (mg)&23.2&24.9&28.8\\
     $\gamma$ (mJ mol$^{-1}$ K$^{-2}$) &4.0&4.02 &3.12\\
     $\beta $ (mJ mol$^{-1}$ K$^{-4}$)&0.12 &0.11&0.10\\
     $\beta_5$ (mJ mol$^{-1}$ K$^{-6}$)&2.48$\times$10$^{-4}$&2.58$\times$10$^{-4}$ & 2.92$\times$10$^{-4}$\\
     $\theta_\mathrm{D}\  (\mathrm{K})$&321&320&323\\
     $D$($\epsilon_\mathrm{F}$) (States/Ry/f.u.) & 23.1 &  23.2  & 18.0 \\
     $\Delta\ (\mathrm{K})$ & 190 &190 &180\\
     $a_1$ (J mol$^{-1}$ K$^{-3/2}$)& 0.31 & 0.40 & 0.32 \\
     \hline
     \hline
     \end{tabular}
\end{table}

Figure \ref{fig5}\textcolor{blue}{d} presents the specific-heat response measured under a 5T magnetic field for different field orientations. To probe the role of spin fluctuations, the field was applied parallel to the [001] and [100] crystallographic axis. Notably, at 5 T, the measurements show no detectable anisotropy among these directions.
The Zeeman energy at 5 T ($\sim$ 0.6 meV $\equiv$ 7 K) is much smaller than the obtained magnon gap (16 meV).
Therefore, a 5 T field does not change the magnon gap or dispersion enough to produce an observable difference in specific heat along different directions.
To understand the variation of $\Delta$ with field, we measured the $C(T)$ under magnetic fields up to $\mu_0H=5$ T as shown in \cref{fig5}\textcolor{blue}{e}.
The specific heat shows a small variation above 100 K (inset of \cref{fig5}\textcolor{blue}{e}).
The field-and temperature-dependent $C(T, H)$ is fitted using \cref{eq7a}, with $\theta_\mathrm{D}$ fixed to the zero-field value, as the phonon contribution is expected to be field-independent. We observe a small reduction of $\Delta$ with increasing field as shown in \cref{fig5}\textcolor{blue}{f}, whereas the coefficient $a_1$ remains nearly constant. This reduction in $\Delta$ is likely due to Zeeman splitting of the lowest-energy magnon mode rather than magnetic anisotropy \cite{vsmejkal2023chiral,singh2025chiral,zhang2025chiral,chen2024magnon}. Nevertheless, due to relatively large uncertainty in $\Delta$, precise discussion on the origin of $\Delta$ change is difficult and urges for in-field spectroscopies.

The magnetic ordering in CrSb is strong, so magnetic excitations are likely to make a significant contribution to the specific heat over a wide temperature range, not just near the N\'{e}el temperature. In collinear antiferromagnets, the absence of magnetic anisotropy preserves continuous spin-rotation symmetry, leading to gapless Goldstone magnon modes, while spin-orbit induced anisotropy opens a finite but degenerate magnon gap \cite{chen2025observation,vsmejkal2023chiral,dos2026comparative}.
In this regard, two prototypical collinear AFMs, NiO and MnO, preserve combined $\mathcal{P}{T}$ symmetry. In the absence of magnetic anisotropy, these systems host degenerate magnon branches. However, the inclusion of weak spin–orbit–induced magnetocrystalline anisotropy lifts this degeneracy near the high symmetry $\Gamma$ point, resulting in a small magnon gap of less than 5 meV \cite{dos2026comparative}.
Altermagnets fundamentally differ from this paradigm \cite{singh2025chiral,vsmejkal2023chiral,zhang2025chiral,singh2025chiral}. In CrSb, the crystal symmetry enforces alternating exchange interactions that lift magnon spin degeneracy even in the collinear and fully compensated state \cite{zhang2025chiral,singh2025chiral}.
Within linear spin-wave theory (LSWT), an effective Heisenberg Hamiltonian with symmetry-inequivalent exchange couplings and single-ion anisotropy supports two spin-split magnon modes. In the long-wavelength limit, the magnon gap \cite{chen2024competing} is
$\Delta_{\rm LSWT} \approx 2S\sqrt{D\,J_{\rm eff}}$, where $S$ is the local moment of the Cr ions, $D$ is the spin-orbit-induced single-ion anisotropy, and $J_{\rm{eff}}=\sum_n z_n J_n$ denotes the effective exchange scale determined by the symmetry-inequivalent Cr-Cr exchange interactions. 
Using reported parameters for CrSb ($S$=3/2, $J_1$= 23 meV, $J_2$= $-5.4$ meV, $J_3$= 5.2 meV, and $D\sim$ 0.15 meV) \cite{singh2025chiral}, the resulting altermagnon gap is estimated to be $\Delta_{\rm LSWT} \sim$ 9-10 meV \cite{zhang2025chiral,singh2025chiral}, consistent with our specific heat fitting.
These suggest that CrSb is a prototypical system where gapped magnons coexist with intrinsic spin-splitting driven by nonrelativistic exchange symmetry.
Based on theoretically predicted non-degenerate magnon dispersion and the experimentally observed large magnon gap supports the realization of the altermagnetic order due to exchange interactions.
\section*{CONCLUSION}
In conclusion, we described details of a high-quality CrSb crystal growth with the self-flux method. 
We obtained large hexagonal facet (001) crystal, up to 2 $\times$ 2.5 $\times$ 1 mm$^3$ with RRR exceeding 10.
A large positive magnetoresistance in CrSb of about 80\%  at 3.5 K and 6 T is observed due to improved sample quality.
Magnetization shows a linear field dependence, consistent with the antiferromagnetic order in CrSb.
The low-temperature specific heat deviates from the conventional $C(T)$ = $\gamma T$ + $\beta T^3$ behavior, which can be well accounted for by higher-order phonon contributions $\beta_5 T^5$.
The extracted Sommerfeld coefficient, $\gamma$ = 4.0 $\pm$ 0.08 mJ mol$^{-1}$ K$^{-2}$, suggests modest electronic correlations among conduction electrons.
At high temperatures, the specific heat surpasses the Dulong–Petit value, which can be attributed to a broad magnon contribution associated with the altermagnetic state. Analysis of the data yields a Debye temperature of 321 $\pm$ 5 K and a magnon excitation gap of approximately 16 $\pm$ 1 meV. Furthermore, no signature of superconductivity is observed in stoichiometric CrSb down to 0.1 K.
Our thermodynamic and transport results advance the understanding of CrSb and establish a robust bases for engineering its altermagnetic order.

\subsection*{ACKNOWLEDGEMENTS}
We are grateful to Yoshiteru Maeno, Soichiro Yamane, Thomas Johnson and Hisakazu Matsuki for useful discussions.
This work was supported by the JSPS KAKENHI (Grants No. JP22H01168, No. JP23K22439, No.~JP26K17093), JST Sakura Science Exchange Program (No. Z2024L1015080) and research support from an IIT Kanpur Initiation Grant (Grant No. IITK-2019-037).
S.P. and C.S. acknowledge research grants from Science and Engineering Research Board (SERB), Government of India (Grants No.SRG2019-001104, No. CRG-2022-005726 and No. EEQ-2022-000883).
G.M. acknowledges support from the Kyoto University Foundation, JSPS KAKENHI (Grant No. JP25K17346), and Toyota Riken Scholar Program.
\\

\subsection*{AUTHOR CONTRIBUTIONS}
\noindent \textbf{Shubhankar Paul:} Conceptualization (lead), Methodology (lead), Data curation (lead), Formal analysis (lead), Visualization (equal), Writing – original draft, Writing – review and editing (equal). 
\textbf{Atsutoshi Ikeda:} Data curation (equal), Writing – review and editing (supporting). 
\textbf{Giordano Mattoni:} Visualization (equal), Data curation (supporting), Writing – review and editing (equal).
\textbf{Shingo Yonezawa:} Conceptualization (equal), Methodology (equal), Visualization (equal), Project administration (equal), Fund acquisition (equal), Writing, review, and editing (equal). 
\textbf{Chanchal Sow:} Conceptualization (equal), Methodology (equal), Visualization (equal), Fund acquisition (equal), Project administration (equal), Writing – review and editing (equal).
\subsection*{ADDITIONAL INFORMATION}
Correspondence and requests for materials should be addressed to  CS and SP.

\subsection*{DATA AVAILABILITY}
The data is available on request to the corresponding author.

\subsection*{REFERENCES}
\bibliography{ref}

@article{Kiefer2025,
  title = {Crystal structure and absence of magnetic order in single-crystalline \ce{RuO2}},
  volume = {37},
  ISSN = {1361-648X},
  DOI = {10.1088/1361-648x/adad2a},
  number = {13},
  journal = {J. Phys. Condens. Matter.},
  publisher = {IOP Publishing},
  author = {Kiefer,  L and Wirth,  F and Bertin,  A and Becker,  P and Bohatý,  L and Schmalzl,  K and Stunault,  A and Rodríguez-Velamazan,  J A and Fabelo,  O and Braden,  M},
  year = {2025},
  month = feb,
  pages = {135801}
}

@article{vsmejkal2022beyond,
  title={Beyond conventional ferromagnetism and antiferromagnetism: A phase with nonrelativistic spin and crystal rotation symmetry},
  author={{\v{S}}mejkal, Libor and Sinova, Jairo and Jungwirth, Tomas},
  journal={Phys. Rev. X},
  volume={12},
  number={3},
  pages={031042},
  year={2022},
  doi={https://doi.org/10.1103/PhysRevX.12.031042},
  publisher={APS}
}

@article{vsmejkal2022emerging,
  title={Emerging research landscape of altermagnetism},
  author={{\v{S}}mejkal, Libor and Sinova, Jairo and Jungwirth, Tomas},
  journal={Phys. Rev. X},
  volume={12},
  number={4},
  pages={040501},
  year={2022},
  doi={https://doi.org/10.1103/PhysRevX.12.040501},
  publisher={APS}
}

@article{karube2022observation,
  title={Observation of spin-splitter torque in collinear antiferromagnetic \ce{RuO2}},
  author={Karube, Shutaro and Tanaka, Takahiro and Sugawara, Daichi and Kadoguchi, Naohiro and Kohda, Makoto and Nitta, Junsaku},
  journal={Phys. Rev. Lett.},
  volume={129},
  number={13},
  pages={137201},
  year={2022},
  doi={https://doi.org/10.1103/PhysRevLett.129.137201},
  publisher={APS}
}

@article{thadathil2026electrical,
  title={Electrical and thermal magnetotransport in altermagnetic \ce{CrSb}},
  author={Thadathil, Sajal Naduvile and M{\"u}ller, Christoph and Firouzmandi, Reza and Farin, Lorenz and Goswami, Srikanta and Badura, Antonin and Manuel, Pascal and Orlandi, Fabio and Ritzinger, Philipp and Pet{\v{r}}{\'\i}{\v{c}}ek, V{\'a}clav and others},
  journal={arXiv preprint arXiv:2603.25820},
  doi={https://doi.org/10.48550/arXiv.2603.25820},
  year={2026}
}

@article{dos2026comparative,
  title={Comparative study of magnetic exchange parameters and magnon dispersions in \ce{NiO} and \ce{MnO} from first principles},
  author={Dos Santos, Flaviano Jose and Binci, Luca and Menichetti, Guido and Mahajan, Ruchika and Marzari, Nicola and Timrov, Iurii},
  journal={Phys. Rev. B},
  volume={113},
  number={2},
  pages={024427},
  doi={10.1103/gtxm-6vtg},
  year={2026},
  publisher={APS}
}

@article{ding2024large,
  title={Large band splitting in g-wave altermagnet \ce{CrSb}},
  author={Ding, Jianyang and Jiang, Zhicheng and Chen, Xiuhua and Tao, Zicheng and Liu, Zhengtai and Li, Tongrui and Liu, Jishan and Sun, Jianping and Cheng, Jinguang and Liu, Jiayu and others},
  journal={Phys. Rev. Lett.},
  volume={133},
  number={20},
  pages={206401},
  doi={10.1103/PhysRevLett.133.206401},
  year={2024},
  publisher={APS}
}

@article{aota2025epitaxial,
  title={Epitaxial growth and transport properties of a metallic altermagnet \ce{CrSb} on a \ce{GaAs} (001) substrate},
  author={Aota, Seiji and Tanaka, Masaaki},
  journal={Phys. Rev. B},
  volume={9},
  number={7},
  pages={074410},
  year={2025},
  doi={10.1103/xjrx-wbv6},
  publisher={APS}
}

@article{bommanaboyena2025single,
  title={Single-crystalline CrSb (0001) thin films grown by dc magnetron co-sputtering},
  author={Bommanaboyena, Satya Prakash and M{\"u}ller, Christoph and Jaro{\v{s}}ov{\'a}, Marketa and Wolk, Katharina and Telkamp, Sjoerd and Zeng, Peng and Krizek, Filip and Uchimura, Tomohiro and Badura, Anton{\'\i}n and Olejn{\'\i}k, Kamil and others},
  journal={Phys. Rev. Mater.},
  volume={9},
  number={6},
    doi={10.1103/PhysRevMaterials.9.064402},
  pages={064402},
  year={2025},
  publisher={American Physical Society}
}

@article{wvs6hqfv,
  title = {Spin-degenerate bulk bands and topological surface states associated with Dirac nodal lines in \ce{RuO2}},
   author={Osumi, T and Yamauchi, K and Souma, S and Shubhankar, P and Honma, A and Nakayama, K and Ozawa, K and Kitamura, M and Horiba, K and Kumigashira, H and others},
  journal = {Phys. Rev. B},
  volume = {113},
  issue = {8},
  pages = {085116},
  numpages = {18},
  year = {2026},
  month = {Feb},
  doi = {10.1103/wvs6-hqfv},
 publisher = {American Physical Society}
}

@article{paul2026multi,
  title={Multi-probe detection of domain nucleation across the metal--insulator transition in \ce{VO2}},
  author={Paul, Shubhankar and Mattoni, Giordano and Ghosh, Amitava and Kesarwani, Pooja and Sahu, Dipak and Ahlawat, Monika and Verma, Amit and Govind Rao, Vishal and Sow, Chanchal and others},
  journal={Appl. Phys. Lett.},
  volume={128},
  number={5},
  pages = {2502226},
  year={2026},
  doi={10.1063/5.0291227},
  publisher={AIP Publishing}
}

@article{Bai2022PRL,
  title = {Observation of Spin Splitting Torque in a Collinear Antiferromagnet},
   author = {Bai,  H. and Han,  L. and Feng,  X.Y. and Zhou,  Y.J. and Su,  R.X. and Wang,  Q. and Liao,  L.Y. and Zhu,  W.X. and Chen,  X.Z. and Pan,  F. and Fan,  X.L. and Song,  C.},
  volume = {128},
  ISSN = {1079-7114},
  DOI = {10.1103/physrevlett.128.197202},
  number = {19},
  journal = {Phys. Rev. Lett.},
  year = {2022},
  pages={197202},
  publisher = {American Physical Society (APS)}
}

@article{paul2025nonanalytic,
  title={Nonanalytic Fermi-liquid correction to the specific heat of \ce{RuO2}},
  author={Paul, Shubhankar and Ikeda, Atsutoshi and Matsuki, Hisakazu and Mattoni, Giordano and Schmalian, J{\"o}rg and Sow, Chanchal and Yonezawa, Shingo and Maeno, Yoshiteru},
  journal={arXiv preprint arXiv:2512.03108},
  doi={https://arxiv.org/abs/2512.03108},
  year={2025}
}

@article{wu2025fermi,
  title={Fermi Surface of \ce{RuO2} Measured by Quantum Oscillations},
  author={Wu, Zheyu and Long, Mengmeng and Chen, Hanyi and Paul, Shubhankar and Matsuki, Hisakazu and Zheliuk, Oleksandr and Zeitler, Uli and Li, Gang and Zhou, Rui and Zhu, Zengwei and others},
  journal={Phys. Rev. X},
  volume={15},
  number={3},
  pages={031044},
  year={2025},
  doi={10.1103/5js8-2hj8},
  publisher={APS}
}

@article{Fedchenko2024SciAdv,
  title={Observation of time-reversal symmetry breaking in the band structure of altermagnetic \ce{RuO2}},
  author={Fedchenko, Olena and Min{\'a}r, Jan and Akashdeep, Akashdeep and D’souza, Sunil Wilfred and Vasilyev, Dmitry and Tkach, Olena and Odenbreit, Lukas and Nguyen, Quynh and Kutnyakhov, Dmytro and Wind, Nils and others},
  journal={Sci. Adv.},
  volume={10},
  number={5},
  pages={eadj4883},
  year={2024},
  doi={10.1126/sciadv.adj4883},
  publisher={American Association for the Advancement of Science}
}

@article{fukami2020antiferromagnetic,
  title={Antiferromagnetic spintronics},
  author={Fukami, Shunsuke and Lorenz, Virginia O and Gomonay, Olena},
  journal={J. Appl. Phys.},
  volume={128},
  number={7},
    pages = {070401},
    month = {08},
  year={2020},
 doi={https://doi.org/10.1063/5.0023614},
  publisher={AIP Publishing}
}

@article{rai2025direction,
  title={Direction-Dependent Conduction Polarity in Altermagnetic \ce{CrSb}},
  author={Rai, Banik and Patra, Krishnendu and Bera, Satyabrata and Kalimuddin, Sk and Deb, Kakan and Mondal, Mintu and Mahadevan, Priya and Kumar, Nitesh},
  journal={Adv. Sci.},
  volume = {12},
 number = {27},
  pages={2502226},
  year={2025},
  doi={10.1002/advs.202502226},
  publisher={Wiley Online Library}
}

@article{jungwirth2016antiferromagnetic,
  title={Antiferromagnetic spintronics},
  author={Jungwirth, Tomas and Marti, X and Wadley, P and Wunderlich, J},
  journal={Nat. Nanotechnol.},
  volume={11},
  number={3},
  pages={231--241},
  year={2016},
doi={10.1038/nnano.2016.18},
  publisher={Nature Publishing Group}
}

@article{paul2025growth,
  title={Growth of ultra-clean oxide single crystals of the altermagnet candidate \ce{RuO2}},
  author={Paul, Shubhankar and Mattoni, Giordano and Matsuki, Hisakazu and Johnson, Thomas and Sow, Chanchal and Yonezawa, Shingo and Maeno, Yoshiteru},
  journal={J. Cryst. Growth},
  volume = {673},
 pages = {128405},
  year={2025},
 doi={10.1016/j.jcrysgro.2025.128405},
  publisher={Elsevier}
}

@article{radhakrishna1996inelastic,
  title={Inelastic-neutron-scattering studies of spin-wave excitations in the pnictides \ce{MnSb} and \ce{CrSb}},
  author={Radhakrishna, P and Cable, JW},
  journal={Phys. Rev. B},
  volume={54},
  number={17},
  pages={11940},
  year={1996},
  doi={https://doi.org/10.1103/PhysRevB.54.11940},
  publisher={APS}
}

@article{li2025large,
  title={Large anomalous Nernst effect in a metallic altermagnet \ce{CrSb} single crystal},
  author={Li, Wei and Xu, Chunqiang and Wang, Minjun and Zou, Mengting and Li, Wan and Wang, Hongwei and Jiang, Wei and Wang, Baomin},
  journal={Phys. Rev. B},
  volume={112},
  number={10},
  pages={L100401},
  year={2025},
doi={ DOI: 10.1103/7f46-nhcg},
  publisher={APS}
}

@article{yuan2020magnetic,
  title={Magnetic structure and uniaxial negative thermal expansion in antiferromagnetic \ce{CrSb}},
  author={Yuan, Jibao and Song, Yuzhu and Xing, Xianran and Chen, Jun},
  journal={Dalton Transactions},
  volume={49},
  number={48},
  pages={17605--17611},
  year={2020},
doi={DOI: 10.1039/D0DT03277H},
  publisher={Royal Society of Chemistry}
}

@article{volny2020electrical,
  title={Electrical transport properties of bulk tetragonal CuMnAs},
  author={Voln{\`y}, J and Wagenknecht, D and {\v{Z}}elezn{\`y}, J and Harcuba, P and Duverger--Nedellec, E and Colman, RH and Kudrnovsk{\`y}, J and Turek, I and Uhl{\'\i}{\v{r}}ov{\'a}, K and V{\`y}born{\`y}, K},
  journal={Phys. Rev. Mater.},
  volume={4},
  number={6},
  pages={064403},
  year={2020},
doi={https://doi.org/10.1103/PhysRevMaterials.4.064403},
  publisher={APS}
}

@article{bai2025nonlinear,
  title={Nonlinear field dependence of Hall effect and high-mobility multi-carrier transport in an altermagnet \ce{CrSb}},
  author={Bai, Yuqing and Xiang, Xinji and Pan, Shuang and Zhang, Shichao and Chen, Haifeng and Chen, Xi and Han, Zhida and Xu, Guizhou and Xu, Feng},
  volume={126},
  number={4},
   pages = {042402},
  year={2025},
 journal={Appl. Phys. Lett.},
  doi={10.1063/5.0240434},
  publisher={AIP Publishing}
}

@article{bodnar2020magnetoresistance,
  title={Magnetoresistance effects in the metallic antiferromagnet \ce{Mn2Au}},
  author={Bodnar, S Yu and Skourski, Y and Gomonay, O and Sinova, J and Kl{\"a}ui, M and Jourdan, M},
  journal={Phys. Rev. Appl.},
  volume={14},
  number={1},
  pages={014004},
  year={2020},
doi={https://doi.org/10.1103/PhysRevApplied.14.014004},
  publisher={APS}
}

@article{singh2025chiral,
  title={Chiral Spin-Split Magnons in the Metallic Altermagnet \ce{CrSb}},
  author={Singh, Ashutosh K and Heinsdorf, Niclas and Mancilla, Abraham A and Bannies, J{\"o}rn and Maity, Avishek and Kolesnikov, Alexander I and Matsuda, Masaaki and Stone, Matthew B and Franz, Marcel and Gaudet, Jonathan and others},
  journal={arXiv preprint arXiv:2511.16086},
  doi={https://doi.org/10.48550/arXiv.2511.16086},
  year={2025}
}

@article{yonezawa2015compact,
  title={Compact AC susceptometer for fast sample characterization down to 0.1 {K}},
  author={Yonezawa, Shingo and Higuchi, Takumi and Sugimoto, Yusuke and Sow, Chanchal and Maeno, Yoshiteru},
  journal={Rev. Sci. Instrum.},
  volume={86},
  number={9},
  pages = {093903},
  year={2015},
 doi={http://dx.doi.org/10.1063/1.4929871 },
  publisher={AIP Publishing}
}

@article{urata2024high,
  title={High mobility charge transport in a multicarrier altermagnet \ce{CrSb}},
  author={Urata, Takahiro and Hattori, Wataru and Ikuta, Hiroshi},
  journal={Phys. Rev. Mater.},
  volume={8},
  number={8},
  pages={084412},
  year={2024},
doi={https://doi.org/10.1103/PhysRevMaterials.8.084412},
  publisher={APS}
}

@article{zhang2025chiral,
  title={Chiral magnon splitting in altermagnetic \ce{CrSb} from first principles},
  author={Zhang, Yi-Fan and Ni, Xiao-Sheng and Chen, Ke and Cao, Kun},
  journal={Phys. Rev. B},
  volume={111},
  number={17},
  pages={174451},
  year={2025},
doi={10.1103/PhysRevB.111.174451},
  publisher={APS}
}

@article{chen2024competing,
  title={Competing itinerant and local spin interactions in kagome metal \ce{FeGe}},
  author={Chen, Lebing and Teng, Xiaokun and Tan, Hengxin and Winn, Barry L and Granroth, Garrett E and Ye, Feng and Yu, DH and Mole, RA and Gao, Bin and Yan, Binghai and others},
  journal={Nat. Commun.},
  volume={15},
  number={1},
  pages={1918},
  year={2024},
doi={https://doi.org/10.1038/s41467-023-44190-2},
  publisher={Nature Publishing Group UK London}
}

@article{corticelli2022spin,
  title={Spin-space groups and magnon band topology},
  author={Corticelli, Alberto and Moessner, Roderich and McClarty, Paul A},
  journal={Phys. Rev. B},
  volume={105},
  number={6},
  pages={064430},
  year={2022},
doi={https://doi.org/10.1103/PhysRevB.105.064430},
  publisher={APS}
}

@article{kahal2007magnetic,
  title={Magnetic properties of \ce{CrSb}: A first-principle study},
  author={Kahal, L and Zaoui, Ali and Ferhat, M},
  journal={J. Appl. Phys.},
  volume = {101},
  number = {9},
  pages = {093912},
  year = {2007},
  month = {05},
  doi={https://doi.org/10.1063/1.2732502},
  publisher={AIP Publishing}
}

@article{dahal2017possible,
  title={Possible superconductivity in chemically doped \ce{CrSb_{1+$\delta$}}},
  author={Dahal, Ashutosh and Gunasekera, Jagath and Singh, Deepak K},
  journal={Phys. Status Solidi RRL.},
  volume={11},
  number={10},
  pages={1700211},
  year={2017},
doi={ 10.1002/pssr.201700211},
  publisher={Wiley Online Library}
}

@article{takei1963magnetic,
  title={Magnetic structures in the \ce{MnSb}-\ce{CrSb} system},
  author={Takei, WJ and Cox, De E and Shirane, G},
  journal={Phys. Rev.},
  volume={129},
  number={5},
  pages={2008},
  year={1963},
doi={10.1103/PhysRev.129.2008},
  publisher={APS}
}

@article{chen2025observation,
  title={Observation of Coherent Gapless Magnons in an Antiferromagnet},
  author={Chen, Jilei and Jin, Zhejunyu and Yuan, Rundong and Wang, Hanchen and Jia, Hao and Wei, Weiwei and Sheng, Lutong and Wang, Jinlong and Zhang, Yuelin and Liu, Song and others},
  journal={Phys. Rev. Lett.},
  volume={134},
  number={5},
  pages={056701},
  year={2025},
doi={DOI: https://doi.org/10.1103/PhysRevLett.134.056701},
  publisher={APS}
}

@article{joshua1998magnon,
  title={Magnon contribution to the specific heat and the validity of power laws in antiferromagnetic crystals},
  author={Joshua, SJ},
  journal={Physica A Stat. Mech. Appl.},
  volume={261},
  number={1-2},
  pages={135--142},
  year={1998},
doi={https://doi.org/10.1016/S0378-4371(98)00370-7},
  publisher={Elsevier}
}

@article{jiang2024enumeration,
  title={Enumeration of spin-space groups: Toward a complete description of symmetries of magnetic orders},
  author={Jiang, Yi and Song, Ziyin and Zhu, Tiannian and Fang, Zhong and Weng, Hongming and Liu, Zheng-Xin and Yang, Jian and Fang, Chen},
  journal={Phys. Rev. X},
  volume={14},
  number={3},
  pages={031039},
  year={2024},
  doi={10.1103/PhysRevX.14.031039},
  publisher={APS}
}

@article{mazin2022altermagnetism,
  title={Altermagnetism—a new punch line of fundamental magnetism},
  author={Mazin, Igor and {PRX editor}},
  journal={Phys. Rev. X},
  volume={12},
  number={4},
  pages={040002},
  doi={10.1103/PhysRevX.12.040002},
  year={2022},
  publisher={APS}
}

@article{peng2025scaling,
  title={Scaling behavior of magnetoresistance and Hall resistivity in the altermagnet \ce{CrSb}},
  author={Peng, Xin and Wang, Yuzhi and Zhang, Shengnan and Zhou, Yi and Sun, Yuran and Su, Yahui and Wu, Chunxiang and Zhou, Tingyu and Liu, Le and Wang, Hangdong and others},
  journal={Phys. Rev. B},
  volume={111},
  number={14},
  pages={144402},
  year={2025},
  doi={10.1103/PhysRevB.111.144402},
  publisher={APS}
}

@article{varshney2004analysis,
  title={Analysis of low temperature specific heat in the ferromagnetic state of the Ca-doped manganites},
  author={Varshney, Dinesh and Kaurav, N},
  journal={Eur. Phys. J. B.},
  volume={37},
  number={3},
  pages={301--309},
  year={2004},
  doi={10.1140/epjb/e2004-00060-x},
  publisher={Springer}
}

@article{yang2025three,
  title={Three-dimensional mapping of the altermagnetic spin splitting in \ce{CrSb}},
  author={Yang, Guowei and Li, Zhanghuan and Yang, Sai and Li, Jiyuan and Zheng, Hao and Zhu, Weifan and Pan, Ze and Xu, Yifu and Cao, Saizheng and Zhao, Wenxuan and others},
  journal={Nat. Commun.},
  volume={16},
  number={1},
  pages={1442},
  year={2025},
  doi={10.1038/s41467-025-56647-7},
  publisher={Nature Publishing Group UK London}
}

@article{chen2024magnon,
  title={Magnon Excitation Modes in Ferromagnetic and Antiferromagnetic Systems},
  author={Chen, Xing and Zheng, Cuixiu and Liu, Yaowen},
  journal={Magnetochemistry},
  volume={10},
  number={7},
  pages={50},
  doi={https://doi.org/10.3390/magnetochemistry10070050},
  year={2024},
  publisher={MDPI}
}

@article{vsmejkal2023chiral,
  title={Chiral magnons in altermagnetic \ce{RuO2}},
  author={{\v{S}}mejkal, Libor and Marmodoro, Alberto and Ahn, Kyo-Hoon and Gonz{\'a}lez-Hern{\'a}ndez, Rafael and Turek, Ilja and Mankovsky, Sergiy and Ebert, Hubert and D’Souza, Sunil W and {\v{S}}ipr, Ond{\v{r}}ej and Sinova, Jairo and others},
  journal={Phys. Rev. Lett.},
  volume={131},
  number={25},
  pages={256703},
  year={2023},
 doi={https://doi.org/10.1103/PhysRevLett.131.256703},
  publisher={APS}
}

@article{liu2024chiral,
  title={Chiral split magnon in altermagnetic \ce{MnTe}},
  author={Liu, Zheyuan and Ozeki, Makoto and Asai, Shinichiro and Itoh, Shinichi and Masuda, Takatsugu},
  journal={Phys. Rev. Lett.},
  volume={133},
  number={15},
  pages={156702},
  year={2024},
  doi={https://doi.org/10.1103/PhysRevLett.133.156702},
  publisher={APS}
}

@article{tas2025magnon,
  title={The magnon spectra of g-type altermagnet bulk \ce{CrSb}},
  author={Tas, Murat},
  journal={arXiv preprint arXiv:2510.03759},
 doi={https://arxiv.org/pdf/2510.03759},
  year={2025}
}

@article{reimers2024direct,
  title={Direct observation of altermagnetic band splitting in \ce{CrSb} thin films},
  author={Reimers, Sonka and Odenbreit, Lukas and {\v{S}}mejkal, Libor and Strocov, Vladimir N and Constantinou, Procopios and Hellenes, Anna B and Jaeschke Ubiergo, Rodrigo and Campos, Warlley H and Bharadwaj, Venkata K and Chakraborty, Atasi and others},
  journal={Nat. Commun.},
  volume={15},
  number={1},
  pages={2116},
  year={2024},
  doi={https://doi.org/10.1038/s41467-024-46476-5},
  publisher={Nature Publishing Group UK London}
}

@article{snow1952neutron,
  title={Neutron diffraction investigation of the atomic magnetic moment orientation in the antiferromagnetic compound \ce{CrSb}},
  author={Snow, AI},
  journal={Phys. Rev.},
  volume={85},
  number={2},
  pages={365},
  year={1952},
doi={https://doi.org/10.1103/PhysRev.85.365},
  publisher={APS}
}

@article{abe1984magnetic,
  title={Magnetic properties of \ce{CrSb}},
  author={Abe, Shunya and Kaneko, Takejiro and Ohashi, Masayoshi and Yoshida, Hajime and Kamigaki, Kazuo},
  journal={J. Phys. Soc. Jpn.},
  volume={53},
  number={8},
  pages={2703--2709},
  year={1984},
  doi={https://doi.org/10.1143/JPSJ.53.2703},
  publisher={The Physical Society of Japan}
}

@article{hirone1956magnetic,
  title={On the magnetic properties of the system \ce{MnSb}--\ce{CrSb}},
  author={Hirone, Tokutaro and Maeda, Seijiro and Tsubokawa, Ichiro and Tsuya, Noboru},
  journal={J. Phys. Soc. Jpn.},
  volume={11},
  number={10},
  pages={1083--1087},
  year={1956},
  doi={https://doi.org/10.1143/JPSJ.11.1083},
  publisher={The Physical Society of Japan}
}

@article{rodriguez1993recent,
  title={Recent advances in magnetic structure determination by neutron powder diffraction},
  author={Rodr{\'\i}guez-Carvajal, Juan},
  journal={Physica B.},
  volume={192},
  number={1-2},
  pages={55--69},
 doi= {https://doi.org/10.1016/0921-4526(93)90108-I},
  year={1993},
  publisher={Elsevier}
}
\bibliographystyle{apsrev4-2}

\end{document}